# DESIGNING A PORTABLE RADIO FREQUENCY SOURCE FOR MAGNETIC HYPERTHERMIA APPLICATIONS

**Christian Alexander Calvache**
Grupo de Investigación en Materiales Funcionales Nanoestructurados,
Universdad CESMAG Pasto,Nariño
calvachechristian@gmail.com

**Jenny Alejandra Mera**
Grupo de Investigación en Materiales Funcionales Nanoestructurados,
Universdad CESMAG Pasto,Nariño
gamera@unicesmag.edu.co

**Diego Fernando Coral**
Departamento de Física, Universidad del Cauca, Popayán-Colombia
dfcoral@unicauca.edu.co

*Abstract*—In the 21st century, new and alternative non-radioactive methods for cancer treatment have been developed. One of this is magnetic hyperthermia (MH), which is based on the ability of magnetic nanoparticles (MNP) to transduce electromagnetic energy into heat. If MNP are inside cancer cells, heat induces cells death. The success of HM as cancer therapy depends on two factors, first, the NPM properties and second, the frequency and amplitude of the radio frequency (RF) applied field. Therefore, in this work a design of a portable radio frequency source for magnetic hyperthermia experiments is presented. The design consists of two parts, the first part corresponds to the RF source composed by a RLC parallel circuit, the capacity and inductance values have been chosen in order to achieve the RF range of frequency, and the second part corresponds to the design of a magnetization sensor circuit. While the first circuit produces the RF field, the second circuit serves as a detector of the NPM induced magnetization. The development of this radiofrequency source makes it possible to determine the applicability of an NPM system in the treatment of cancer by magnetic hyperthermia, opening doors for future research focused on the application of nanotechnology in biomedicine.

## I. INTRODUCTION

The most common methods used for the treatment of cancer are chemotherapy and radiotherapy, but in the XXI century with new investigations, non-radioactive alternative methods such as magnetic hyperthermia (MH) [1] have been developed. In MH magnetic nanoparticles (MNP) composed by magnetic material (Fe, Ni, Co) and coated with a biocompatible layer are used to transform the electromagnetic field energy into heat [2]. If MNP are inside the tumor tissue, this can increase the cells temperature until activate the cell death process known as apoptosis. This process becomes active when the tumor temperature rises between 42-45 °C [3].

For MH applications, a radiofrequency (RF) source is needed in order to provide magnetic energy to the MNP. This source is composed by an RLC circuit, which, with the appropriate parameters, can produce alternate electromagnetic fields in the range of 100 kHz to 150 kHz with field amplitudes between 2 kA/m and 15 kA/m [4].

In this work, a design of a radio frequency source is presented. The source consists of a parallel RLC, circuit working in resonance, which generates an electromagnetic field, and a sensor circuit which will be used to measure the magnetic response of a group of nanoparticles submitted to RF fields. For this, as first step, a field-generator coil and a sensor coil have been designed, the inductance (L) of the coils was determined measuring the resonance frequency using different capacitances values. In a second part, an AC/DC and DC/AC converters simulation is presented in order to fit the circuit parameters and to achieve the power required by the circuit.

When MNP are immersed in a magnetic field $H$, the energy is used to alienate the magnetic moments in the same field direction. The volume density of magnetic moments aligned with the field is called magnetization ($M$). For MH applications, both, $H$ amplitude and $M$ vary sinusoidally, but the magnetization response is delayed in respect to $H$ due to magnetic relaxation processes [5]. This delay is the responsible of heating.

By this way, the widest the magnetization cycle, the higher the delay time and the higher the heating [6]. The last means that the applicability of a MNP set in MH can be studied by analyzing the MNP magnetization cycle, for this, in this work a magnetization sensing circuit is also presented as a complementary part of the RF source for HM applications.

For magnetization sensing, a coil and an amplifier circuit are used. On this coil an electromotive force is induced due to the time-depended sample magnetization produced by the RF field, the induced signal has the same frequency that the applied RF field and the current intensity is directly proportional to the sample magnetization [7].

By this way, it is expected that the obtained results can be useful in the fabrication of RF sources. On the other hand, It is expected that from this designs a work-instrument portable and functional in different measurement environments such as laboratories and hospitals can be manufactured.

## II. METHODOLOGY

### A. Inductance Determination

The main component of a RF source is a resonant RLC circuit. A capacitance and a field-generator coil compose it. In order to study the coil size effects, two field-generator coils were made. The inductance value of was determined measuring the resonance frequency (ω) at different capacitance values. In the same way other set of sensing-coils were built and the inductance was also determined using the same method.

The main coil characteristics are listed next:

- Generator coil α: 36 turns copper wire caliper 20 winding with PVC cover. This coil has a diameter of 3,37±0,01 cm and a length of 10,12±0,01 cm.
- Generator coil β: 41 turns copper wire caliper 20 winding with PVC cover. This coil has a diameter of 4,91±0,01 cm and a length of 10,87±0,01 cm.
- Sensor coil: 10 turns enameled copper wire caliper 18 winding. This coil has a diameter of 1,15±0,01 cm and a length of 1±0,01 cm.

To determine the inductance of each coil, an LC circuit as showed in figure 1(a) was implemented. The capacitances were varied between 1 μF and 220 μF and during this process the frequency was adjusted to reach the resonance state correspondent to each value of C. This process was also carried out to determine L of the sensor coil.

Once connected the LC circuit, the frequency was varied until reach the resonance state, in this state, the frequency is given by the equation (1)

$$\omega = \frac{1}{\sqrt{LC}} \qquad (1)$$

Being $\omega = 2\pi f$, and $f$ the generator frequency.

The data obtained in the experiments were used to realize a chart of $1/\omega^2$ as a function of C, the slope of this curve represents the inductance (L) of the used coil.

### B. Simulation of the AC/DC and DC/AC converter

For the conversers system simulation the Proteus 8.7 Software was used. The simulated AC/DC converser circuit consists of a 110V at 60 Hz input, 4 rectifying bridges, capacitors of 10pF and 2200uF to filter the signal, a voltage regulator 7805. The DC/AC converter circuit consists of four optocouplers 6N137 for the PWM input, resistances, 2N3904 NPN transistors to amplify the signal, protection diodes and four MOSFET transistors, which are responsible for commute and amplify the signal, with the final purpose of obtain a frequency superior to 10 kHz.

## III. RESULTS

The circuit block diagrams of the presented design is shown in Fig. 1.

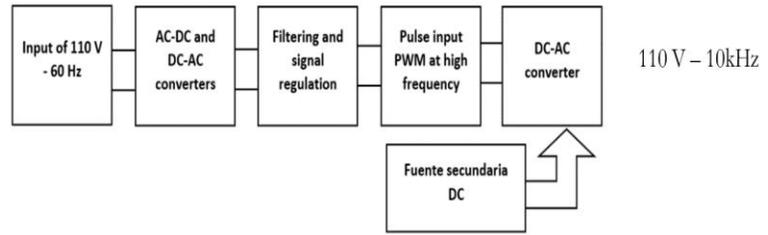

Fig. 1. Converters blocks diagram.

The source consists in a LC circuit working at resonance in the RF range. By this way, the main component of the RF source is the field-generator coil. A variation on coils parameters is observed as a variation in the resonance frequency as expressed by equation 1. In order to obtain data about the length and shape of the main inductor, two types of field-generator coils (α and β) were studied.

For inductance determination, and LC circuit were used (as shown in Fig. 2) and its resonance state was studied, it is observed as a close loop figure in an oscilloscope using the XY visualization mode, in which, the input signal was measured in channel 1 and the inductance signal on the channel 2. Fig. 2 shows a representative photograph of circuit signal in the resonance state.

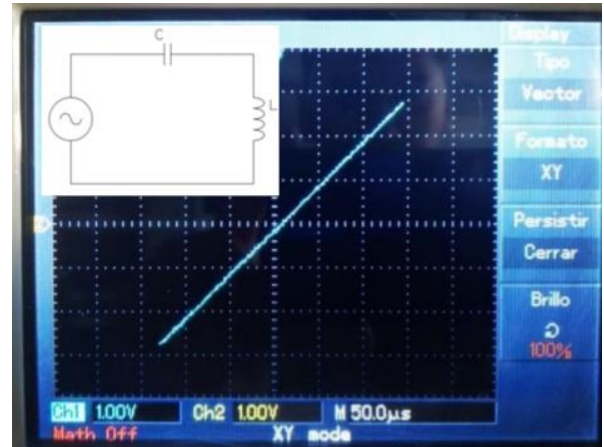

Fig. 2. Oscilloscope screen photograph showing the difference between the input signal and the inductance signal in the resonance state. *Inset:* Scheme of the implemented circuit to determine the coils inductance.

Once built the LC circuit, the inductance of each field-generator coil was determined using different values of capacitance. Fig. 3 shows the behavior of $1/\omega^2$ as function of capacitance for all the studied coils.

The observed linear behavior in Figure 4, validates the use of equation 1 to obtain the inductance value of each coil using a linear fit. The L values are presented in table 1.

Fig. 1. Inverse square resonance frequency ($1/\omega^2$) as function of capacity for (a) coil α, (b) coil β and (c) sensor coil. Solid line corresponds linear fit.

Once determined the field-generator coil inductance, as third step, the AC/DC and a DC/AC convertor were simulated using Proteus software. The converters were built using a 110V at 60 Hz input, four full wave rectifying bridges which fulfill the function of invert the negative voltage resulting in a continuous signal. The signal passes through a filter sequence and a voltage regulator to stabilize the system. Then, the signal is taken to an optocoupler, which is responsible for input of the (pulse-width modulation) PWM pulses, which is previously programmed in Arduino, and the protection of the converter.

Through NPN transistors the signal will be amplify, increasing the efficiency of the system. That signal is sent to the MOSFETs that are the responsible to commute and amplify the frequency signal. The assembly of the AC-DC and DC-AC convertors circuit are presented in Fig. 4 and 5.

Table 1 Inductance values obtained for each coil.

As second, an important component annexed to the RF source is the sensing system. This system is designed in order to sense the sample magnetic response to RF fields. For this, a coil named sensor was built and its inductance was calculated using the same procedure used above.

Fig. 3(c) shows the linear square-inverse resonance frequency behavior as function of capacitance for the sensor coil and the obtained *L* value is presented in Table 1.

With the data capture of the prototype, the implementation of a converter system is required, which will give the necessary frequency to carry out the implementation of the radio frequency source.

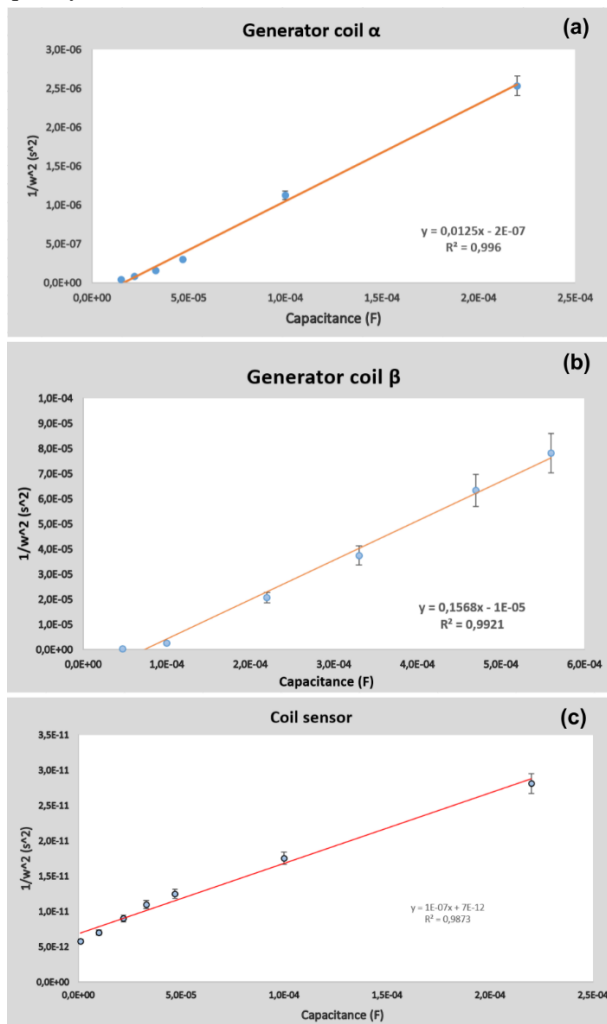

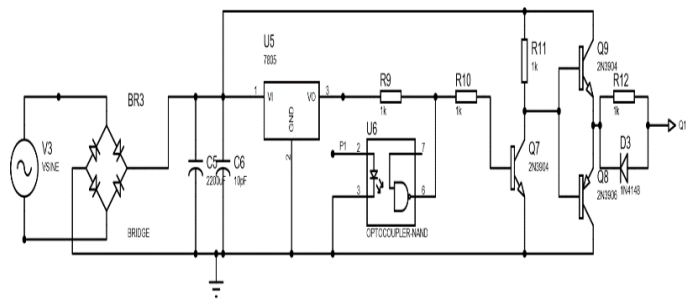

Fig. 4. AC-DC converter.

| Coil | L(H) | Turns |
|---|---|---|
| α | 0.0125 ± 0.004 | 36 |
| β | 0.157 ± 0.008 | 41 |
| Sensor | (1.00±0.01) $x10^{-7}$ | 10 |

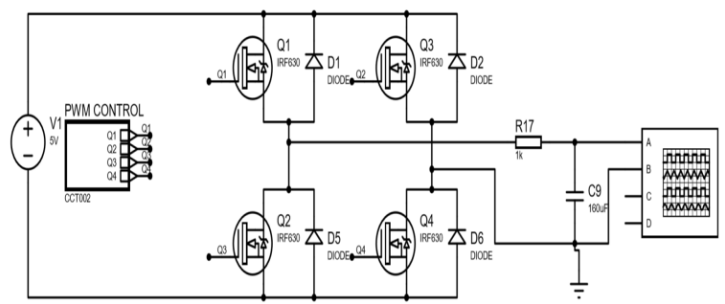

Fig. 5. High DC-AC converter.

The performed circuit simulations gave information about the digital behavior of the system, so that it is possible to implement the circuit and vary the conditions of the same from the simulation.

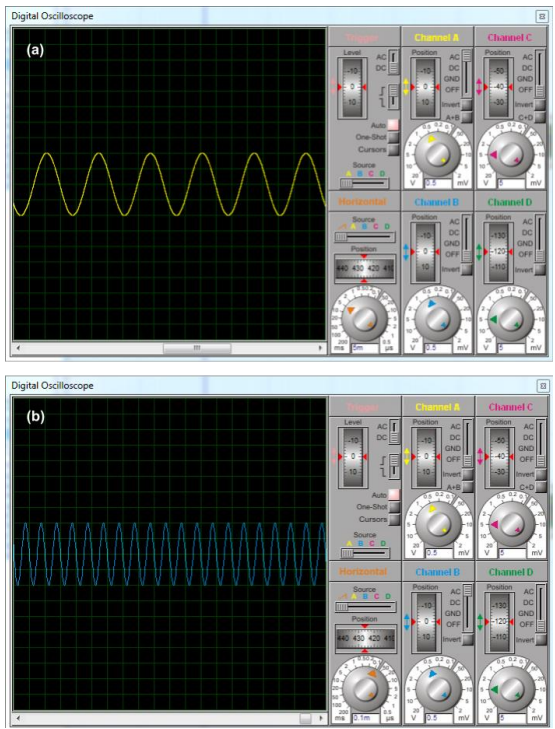

Fig. 6. (a). Voltage (V) vs times(s) of the signal input at 60 Hz. Each horizontal division corresponds to 5 ms. (b). Voltage (V) and times(s) of the output signal at 10 kHz. Each horizontal division corresponds to 0.1 ms.

As result, in figure 6 (a) the input signal at 110 V and 60 Hz is presented. Once the signal passes through the convertors, the frequency is raised until 10 kHz keeping the same voltage amplitude (figure 8 (b)). In each channel, the amplitude of the signal can be graduated, besides the time of the signals can be varied in the horizontal box to be able to visualize a cleaner signal.

## CONCLUSIONS

The results obtained from this work are useful to design a low cost and portable radio frequency source for magnetic hyperthermia applications. From the inductance measurements and the simulations, adequate parameters for circuits were established. From the measurements of resonance frequency as a function of the capacitance, it was possible to determine the inductance field-generator and sensor coils. Likewise, the simulation of the AC / DC and DC / AC converter circuits allows implementing changes in the circuit prior to the implementation of the RF source.